\documentclass{article}
\usepackage{amssymb}
\usepackage{amsmath}

\input{tcilatex}

\begin{document}

\begin{center}
\bigskip {\LARGE Fields tell matter how to move }

Trevor W. Marshall

CCAB, Cardiff University, UK

marshallt@Cardiff.ac.uk
\end{center}

\textbf{Abstract. }Starting from the Oppenheimer-Snyder solution for
gravitational collapse, we show by putting it into the harmonic coordinates,
for which the distant Riemann metric is galilean, that the final state of
collapse for a collapsed star of any mass, including the one thought to
occupy the centre of our galaxy, has a finite radius roughly equal to its
Schwarzschild radius. By applying an expression for the gravitational energy
tensor, we are able to explain the concentration of stellar material in a
thin shell close to the surface, which gives an explanation for why such a
star does not undergo further collapse to a black hole. The interior of the
star is characterized by a low density of the original stellar material,
but, far from being empty, this region is occupied by a very high density of
gravitational energy; this density is negative and the consequent repulsion
is what produces the surface concentration of stellar material.

\section{\protect\bigskip Introduction}

There is a widespread belief that space tells matter how to move\cite%
{wheeler}. This, we believe, has resulted in a profound misunderstanding of
gravity. Although the error can be traced back all the way to the founder of
the modern theory\cite{Einstein1}, \ there is ample evidence that he
nevertheless had a strong inclination to go in the direction which we are
advocating here.

Throughout the decade in which Einstein discovered what he called General
Relativity (GR), he repeatedly attempted to cast gravity within the context
of a classical field theory by constructing a tensor for the energy and
momentum carried by its field\cite{Einstein2}. Perhaps the culmination of
this effort was the article\cite{Einstein3} \ in which he derived the
quadrupole formula for the emission of gravitational radiation from a
bounded nonspherical source. He had already changed his mind twice about
such radiation when he wrote that article; throughout the subsequent six
decades, not only would he himself change two more times, but his
oscillations of opinion would be reflected in the communal understanding of
gravity, causing confusion among virtually all of the leading scholars in
the field\cite{Kennefick}. The communal view changed radically when
observation of the Hulse-Taylor pulsar\cite{HulseT} confirmed that this
system is losing energy at the rate predicted by Einstein's quadrupole
formula.....

But did it? In our opinion the change is incomplete and inadequate. Probably
most of us now confidently expect that the LIGO experiment\cite{LISA} will
reveal that gravitational waves passing through space cause a change in the
path of light signals exchanged across that same space. The light signals
are items of "matter", which is being moved but by what? We have become lazy
in repeating, as a kind of mantra, that the moving agent is space itself.
That may be an adequate description of the static solar field's effect on
planets and passing light signals, but the time has now come to distinguish
between the gravitational field and the space carrying it. The field has
energy which may be localized in a determinate manner; the energy has mass
which itself gravitates, and above all it must be described by an energy
tensor.

In spite of appearing in Einstein's own articles, both before and after GR
was conceived\cite{Einstein2}, the energy "pseudotensor" has not achieved
full tensorial status right up to the present day. There is a big obstacle
to be overcome, and it was pointed out by Hilbert\cite{Hilbert}, who was the
codiscoverer of the basic Hilbert-Einstein field equation of GR. This is
that the metric and its curvature are \emph{the only proper tensors} which
may be constructed out of $g_{\mu \nu }$. The analogy with classical field
theories suggests that an energy tensor be formed from the derivatives of $%
g_{\mu \nu },$ but normal derivation is not tensorial and covariant
derivation in the Riemann metric gives zero.

It must have been physical intuition which caused Einstein to ignore
Hilbert's objection and continue with his "pseudotensor" to deduce the
quadrupole formula. In order to make his argument work he made a particular
coordinate choice, using what is now called the \emph{harmonic} system,
thereby violating the Principle of Equivalence (PE), which he had put at the
centre of his derivation of the Hilbert-Einstein equation, and which states
that no particular coordinate system is to be preferred. Now we are able to
see that, rather than introducing a favoured frame, what he was actually
doing was to recognize that the gravitational field is carried by the
familiar Minkowski space of so called "Special" relativity.\emph{\ }The
notion of the gravitational field as being like the electromagnetic field of
Faraday and Maxwell was taken up by Einstein's contemporaries de Donder and
Lanczos\cite{dedond}\cite{lanczos}, and subsequently developed by Fock\cite%
{fock}, Rosen\cite{rosen} and Weinberg\cite{weinberg}. This field
interpretation of gravity has subsequently been developed, by Logunov and
Mestvirishvili\cite{logmest}, and by Babak and Grishchuk\cite{babak}, to the
point where the Minkowski metric is explicitly present in the field
equations.

In advocating the necessity for including the Minkowski metric in the field
equations, we will also be stressing the need to distinguish between the
various requirements which have been laid on the theories of gravitation,
namely covariance, gauge invariance and the Principle of Equivalence. We
shall argue for maintaining the first, abandoning of the second, and
accepting only the weakest form of the third, in the form of the E\"{o}tv%
\"{o}s Principle (EoP).

Our new contribution to the field theory of gravity is in gravitational
collapse. Starting from the Oppenheimer-Snyder solution\cite{oppsny}, we
show by putting it into the harmonic coordinates, for which the distant
Riemann metric is galilean, that the final state of collapse, for a star of
any mass, including the one thought to occupy the centre of our galaxy, has
a finite radius roughly equal to its Schwarzschild radius.

\section{The energy tensor}

We follow Babak and Grishchuk\cite{babak} (BG) with some changes of
notation. The self-interacting tensor field $h^{\alpha \beta }\left(
x\right) $ is defined on an underlying (flat) Minkowski space with metric%
\begin{equation}
d\sigma ^{2}=\gamma _{\mu \nu }\left( x\right) dx_{\mu }dx_{\nu }\quad .
\end{equation}%
The d'Alembertian operator is%
\begin{equation}
\square _{\gamma }=\gamma ^{\mu \nu }D_{\mu }D_{\nu }\quad ,  \label{minkdal}
\end{equation}%
where 
\begin{equation}
\gamma ^{\mu \nu }\gamma _{\nu \lambda }=\delta _{\lambda }^{\mu }\quad ,
\end{equation}%
and $D_{\mu }$ denotes covariant differentiation with respect to $x_{\mu }$
in the Minkowski metric. We define%
\begin{equation}
\Phi ^{\alpha \beta }=\gamma ^{\alpha \beta }+h^{\alpha \beta }\quad ,
\end{equation}%
and we shall denote covariant differentiation of $\Phi $ simply by a lower
index, that is%
\begin{equation}
\Phi _{\sigma }^{\alpha \beta }=D_{\sigma }\Phi ^{\alpha \beta }=\gamma
_{;\sigma }^{\alpha \beta }+h_{;\sigma }^{\alpha \beta }=h_{;\sigma
}^{\alpha \beta }\quad .
\end{equation}%
We define also the inverse field tensor $\Psi _{\alpha \beta }$ by%
\begin{equation}
\Psi _{\alpha \beta }\Phi ^{\beta \gamma }=\delta _{\alpha }^{\gamma }\quad .
\end{equation}

The field equations are determined from a lagrangian density%
\begin{equation}
L=L_{g}+L_{m}\quad .
\end{equation}%
The gravitational lagrangian is%
\begin{equation}
L_{g}=-\frac{\sqrt{-\gamma }}{4\kappa }\Phi _{\alpha }^{\beta \gamma
}P_{\beta \gamma }^{\alpha },\quad \gamma =\text{det}\left( \gamma _{\mu \nu
}\right) ,\quad \kappa =\frac{8\pi G}{c^{4}},
\end{equation}%
where%
\begin{equation}
P_{\beta \gamma }^{\alpha }=\frac{1}{4}\sqrt{\frac{\gamma }{\Phi }}\left[
2\Phi _{\tau }^{\sigma \tau }\left( \delta _{\beta }^{\alpha }\Psi _{\gamma
\sigma }+\delta _{\gamma }^{\alpha }\Psi _{\beta \sigma }\right) +\Phi
_{\tau }^{\rho \mu }\Phi ^{\tau \alpha }\left( \Psi _{\beta \gamma }\Psi
_{\rho \mu }-2\Psi _{\beta \rho }\Psi _{\gamma \mu }\right) \right] \quad ,
\end{equation}%
and%
\begin{equation}
\quad \Phi =\text{det}\left( \Phi ^{\alpha \beta }\right) \quad .
\end{equation}%
This is equivalent to an expression first given by Fock (Ref\cite{fock},
Appendix B). The "matter" lagrangian $L_{m}$ is a function of all the
nongravitational (particle and field) quantities, collectively labelled by $%
\phi _{m}$, and of $h^{\alpha \beta }$, which enters only through the
combination $\sqrt{-\gamma }\Phi ^{\alpha \beta }$, that is%
\begin{equation}
L_{m}=L_{m}\left( \phi _{m},\sqrt{-\gamma }\Phi ^{\alpha \beta }\right)
\quad .
\end{equation}%
The field equations are then obtained by varying $L$ with respect to $%
h^{\alpha \beta }$, or equivalently $\Phi ^{\alpha \beta }$, and are%
\begin{equation}
-\frac{2\kappa }{\sqrt{-\gamma }}\frac{\delta L_{g}}{\delta \Phi ^{\mu \nu }}%
=-D_{\alpha }P_{\mu \nu }^{\alpha }-P_{\mu \beta }^{\alpha }P_{\nu \alpha
}^{\beta }+\frac{1}{3}P_{\mu \alpha }^{\alpha }P_{\nu \beta }^{\beta }=\frac{%
2\kappa }{\sqrt{-\gamma }}\frac{\partial L_{m}}{\partial \Phi ^{\mu \nu }}%
\quad .  \label{HEBG}
\end{equation}%
BG showed that these are equivalent to the standard Hilbert-Einstein (HE)
equations of GR. First one makes the identification of the inverse field
tensor with the Riemannian metric tensor, namely 
\begin{equation}
g_{\alpha \beta }=\sqrt{\frac{\Phi }{\gamma }}\Psi _{\alpha \beta },\quad
\gamma =\text{det}\left( \gamma ^{\alpha \beta }\right) \quad ,
\label{riemet}
\end{equation}%
from which it follows that%
\begin{equation}
\text{det}\left( g_{\alpha \beta }\right) =g=\Phi /\gamma ^{2}\quad .
\end{equation}%
Then, in terms of the new variables, one finds that 
\begin{equation}
-\frac{2\kappa }{\sqrt{-\gamma }}\frac{\delta L_{g}}{\delta \Phi ^{\mu \nu }}%
=R_{\mu \nu }\quad ,
\end{equation}%
the right side being the contracted curvature tensor of the Riemannian
metric. On the other hand, defining the inverse metric tensor $G^{\mu \nu }$%
(conventional notation $g^{\mu \nu }$) by%
\begin{equation}
G^{\mu \nu }g_{\nu \lambda }=\delta _{\lambda }^{\mu }\quad ,
\end{equation}%
and the material stress tensor by%
\begin{equation}
T_{\mu \nu }=\frac{2c^{2}}{\sqrt{-g}}\frac{\partial L_{m}}{\partial G^{\mu
\nu }}\quad ,
\end{equation}%
we obtain, since $G^{\mu \nu }=\sqrt{\gamma /\Phi }\Phi ^{\mu \nu }$,%
\begin{equation}
\frac{2c^{2}}{\sqrt{-\gamma }}\frac{\partial L_{m}}{\partial \Phi ^{\mu \nu }%
}=T_{\mu \nu }-\frac{1}{2}g_{\mu \nu }G^{\alpha \beta }T_{\alpha \beta
}=T_{\mu \nu }-\frac{1}{2}g_{\mu \nu }T\quad .
\end{equation}%
With these substitutions (\ref{HEBG}) becomes%
\begin{equation}
R_{\mu \nu }=\kappa c^{2}\left( T_{\mu \nu }-\frac{1}{2}g_{\mu \nu }T\right)
\quad ,  \label{hilbein}
\end{equation}%
which is the Hilbert-Einstein equation\footnote{%
Our derivation follows a long line of field theoretic ones, starting with
Hilbert, who used the lagrangian $R\sqrt{-g}.$ The lagrangian of BG is a
covariant version of one used from the years shortly after the establishing
of GR (see, for example, Eddington\cite{Eddington} section 58).}.

The energy balance is obtained by varying $L$ with respect to $\gamma _{\mu
\nu }$, that is (note that this variation takes account of $\gamma ^{\mu \nu
}$ occurring in $\Phi ^{\mu \nu }$), 
\begin{equation}
\frac{\delta L_{g}}{\delta \gamma _{\mu \nu }}+\frac{\partial L_{m}}{%
\partial \gamma _{\mu \nu }}=0\quad .  \label{ebalance}
\end{equation}%
The first term is given by%
\begin{equation}
-\frac{2}{\sqrt{-\gamma }}\frac{\delta L_{g}}{\delta \gamma _{\mu \nu }}=-%
\frac{1}{2\kappa }D_{\alpha }D_{\beta }\left( \Phi ^{\mu \nu }\Phi ^{\alpha
\beta }-\Phi ^{\mu \alpha }\Phi ^{\nu \beta }\right) -\frac{2}{\sqrt{-\gamma 
}}q^{\alpha \beta \mu \nu }\frac{\delta L_{g}}{\delta \Phi ^{\alpha \beta }}%
+t_{g}^{\mu \nu }\quad ,
\end{equation}%
where%
\begin{equation}
q^{\alpha \beta \mu \nu }=\Phi ^{\alpha \mu }\Phi ^{\beta \nu }-\frac{1}{2}%
\Phi ^{\alpha \beta }\Phi ^{\mu \nu }-\gamma ^{\alpha \mu }\gamma ^{\beta
\nu }+\frac{1}{2}\gamma ^{\mu \nu }\Phi ^{\alpha \beta }\quad ,
\end{equation}%
and\footnote{%
This expression was given in noncovariant form, that is with ordinary
instead of covariant derivatives, by Landau and Lifshitz\cite{LL} and also
by Fock\cite{fock}, as the gravitational energy "pseudotensor".}%
\begin{eqnarray}
16\kappa t_{g}^{\mu \nu } &=&\left( 2\Phi ^{\mu \delta }\Phi ^{\nu \omega
}-\Phi ^{\mu \nu }\Phi ^{\delta \omega }\right) \left( 2\Psi _{\alpha \rho
}\Psi _{\beta \sigma }-\Psi _{\alpha \beta }\Psi _{\rho \sigma }\right) \Phi
_{\delta }^{\rho \sigma }\Phi _{\omega }^{\alpha \beta }  \notag \\
&&+8\Phi ^{\rho \sigma }\Psi _{\alpha \beta }\Phi _{\sigma }^{\nu \beta
}\Phi _{\rho }^{\mu \alpha }-8\Phi ^{\mu \alpha }\Psi _{\beta \rho }\Phi
_{\sigma }^{\nu \beta }\Phi _{\alpha }^{\rho \sigma }-8\Phi ^{\nu \alpha
}\Psi _{\beta \rho }\Phi _{\sigma }^{\mu \beta }\Phi _{\alpha }^{\rho \sigma
}  \notag \\
&&+4\Phi ^{\mu \nu }\Psi _{\alpha \rho }\Phi _{\sigma }^{\alpha \beta }\Phi
_{\beta }^{\rho \sigma }+8\Phi _{\rho }^{\mu \nu }\Phi _{\sigma }^{\rho
\sigma }-8\Phi _{\alpha }^{\mu \alpha }\Phi _{\beta }^{\nu \beta }\quad .
\end{eqnarray}%
Hence, making use of the field equation (\ref{HEBG}), this first term becomes%
\begin{equation}
-\frac{2}{\sqrt{-\gamma }}\frac{\delta L_{g}}{\delta \gamma _{\mu \nu }}=-%
\frac{1}{2\kappa }D_{\alpha }D_{\beta }\left( \Phi ^{\mu \nu }\Phi ^{\alpha
\beta }-\Phi ^{\mu \alpha }\Phi ^{\nu \beta }\right) +\frac{2}{\sqrt{-\gamma 
}}q^{\alpha \beta \mu \nu }\frac{\partial L_{m}}{\partial \Phi ^{\alpha
\beta }}+t_{g}^{\mu \nu }\quad .  \label{lgderiv}
\end{equation}%
The second term is given by%
\begin{equation}
\frac{\partial L_{m}}{\partial \gamma _{\mu \nu }}=\left( \gamma ^{\mu
\alpha }\gamma ^{\nu \beta }-\frac{1}{2}\gamma ^{\mu \nu }\Phi ^{\alpha
\beta }\right) \frac{\partial L_{m}}{\partial \Phi ^{\alpha \beta }}\quad .
\label{lmderiv}
\end{equation}%
Hence (\ref{ebalance}) becomes%
\begin{gather}
\frac{1}{2\kappa }D_{\alpha }D_{\beta }\left( \Phi ^{\mu \nu }\Phi ^{\alpha
\beta }-\Phi ^{\mu \alpha }\Phi ^{\nu \beta }\right) =t_{g}^{\mu \nu }-\frac{%
2}{\sqrt{-\gamma }}\left( \Phi ^{\alpha \mu }\Phi ^{\beta \nu }-\frac{1}{2}%
\Phi ^{\alpha \beta }\Phi ^{\mu \nu }\right) \frac{\partial L_{m}}{\partial
\Phi ^{\alpha \beta }}  \notag \\
=t_{g}^{\mu \nu }+\frac{g}{\gamma }\left( G^{\alpha \mu }G^{\beta \nu }-%
\frac{1}{2}G^{\alpha \beta }G^{\mu \nu }\right) \left( T_{\mu \nu }-\frac{1}{%
2}g_{\mu \nu }G^{\sigma \tau }T_{\sigma \tau }\right) \quad .
\end{gather}%
In the usual notation the inverse Riemann metric is $G^{\alpha \beta
}=g^{\alpha \beta }$, and it is the tensor used for raising indices, so that
the right side simplifies to give%
\begin{equation}
D_{\alpha }D_{\beta }\left( \Phi ^{\mu \nu }\Phi ^{\alpha \beta }-\Phi ^{\mu
\alpha }\Phi ^{\nu \beta }\right) =2\kappa t^{\mu \nu },\quad t^{\mu \nu
}=\left( t_{g}^{\mu \nu }+\frac{g}{\gamma }T^{\mu \nu }\right) \quad .
\label{ebalance1}
\end{equation}%
Now the expression $D_{\alpha }D_{\beta }D_{\mu }\left( \Phi ^{\mu \nu }\Phi
^{\alpha \beta }-\Phi ^{\mu \alpha }\Phi ^{\nu \beta }\right) $ is
antisymmetric with respect to a $\mu \alpha $ interchange, from which we may
deduce the covariant conservation equation%
\begin{equation}
D_{\mu }t^{\mu \nu }=0\quad .
\end{equation}%
The theory summarized in this section has all three basic properties
mentioned in the previous section, that is covariance, gauge invariance and
weak equivalence as summed up in EoP. The latter may be briefly stated as
"All forms of energy gravitate equally". It has an active aspect
characterized by the single coupling constant $\kappa $ in (\ref{ebalance1}%
), and also a passive aspect, contained in the equation%
\begin{equation}
\nabla _{\mu }T^{\mu \nu }=0\quad .  \label{EOPpass}
\end{equation}%
Of course, the latter property is a consequence of (\ref{hilbein}) together
with the Bianchi identity%
\begin{equation}
\nabla _{\mu }\left( R^{\mu \nu }-\frac{1}{2}g^{\mu \nu }R\right) =0\quad .
\end{equation}

\section{The harmonic condition}

The condition%
\begin{equation}
\partial _{\mu }\Phi ^{\mu \nu }=0  \label{harmonic}
\end{equation}%
has a history almost as long as GR itself. It was used by Einstein\cite%
{Einstein3} in his derivation of the quadrupole formula, but was almost
immediately criticized, for example by Eddington (Ref\cite{Eddington}, page
130), for being noncovariant; Eddington said that gravitational waves
"travel at the speed of thought", and his criticism resounded throughout the
subsequent six decades, finding fame in the title of a seminal history\cite%
{Kennefick} of gravitational waves. In the first decade of GR the harmonic
condition was developed by de Donder\cite{dedond}, and also by Einstein's
research assistant Lanczos\cite{lanczos}; later it was championed by Fock%
\cite{fock}, who claimed that this choice of coordinates was necessary in
order to guarantee the condition that there are no ingoing waves at large
distance from an isolated massive object. Fock, and also Weinberg\cite%
{weinberg} have shown that this condition, as well as being an essential
part of the field interpretation of GR, is extremely practical in all
analyses of the far fields of such a system, for example in establishing
post-Newtonian expansions. Subsequently\cite{logunov} it was argued by
Logunov and Mestvirishvili that, in its covariant form%
\begin{equation}
D_{\mu }\Phi ^{\mu \nu }=0\quad ,  \label{coharmonic}
\end{equation}%
the harmonic condition, far from being just a choice of coordinate system or
"gauge", is, along with the Hilbert-Einstein equation dealt with in the
previous section, an essential field equation of gravitation. These authors
showed that, by including the Minkowski metric $\gamma _{\mu \nu }$
explicitly in a lagrangian%
\begin{equation}
L_{g1}=\lambda \left( 2\sqrt{-\gamma }\gamma _{\mu \nu }\Phi ^{\mu \nu }-%
\sqrt{-g}\right) \quad ,
\end{equation}%
so that the total lagrangian is%
\begin{equation}
L=L_{g}+L_{g1}+L_{m}\quad ,
\end{equation}%
the covariant harmonic condition does indeed appear as a field equation.
Such a modification causes the Hilbert-Einstein equation to acquire two
additional terms, one of which is the familiar "cosmological constant" term
(what Einstein termed his "biggest mistake", which, however, has now come
back into fashion). The cosmological effect of these two terms has been
explored\cite{logunov} with $\lambda $ related to the Hubble constant, but
in this article we shall assume that $\lambda $ is negligible, so that the
Hilbert-Einstein equation remains unmodified. The extra term $\delta
L_{g1}/\delta h^{\mu \nu }$ in the field equation violates the gauge
invariance, though not, we stress, the covariance. In general this
additional term may result also in the violation of EoP, in the form (\ref%
{EOPpass}), but the imposition of the field equation (\ref{coharmonic}) is
precisely what prevents this occurrence. It should be noted that the loss of
gauge invariance arises precisely because of this additional field equation,
and that in field theoretic terms it is more accurate to state that what has
been lost is the \emph{gauge ambiguity }of GR.

As far as the energy tensor is concerned, we may incorporate the new
equation and write the energy balance (\ref{ebalance1}) as%
\begin{equation}
t_{H}^{\mu \nu }=t_{gH}^{\mu \nu }+\frac{g}{\gamma }T^{\mu \nu }\quad ,
\label{ebalharm}
\end{equation}%
where%
\begin{eqnarray}
16\kappa t_{gH}^{\mu \nu } &=&\left( 2\Phi ^{\mu \delta }\Phi ^{\nu \omega
}-\Phi ^{\mu \nu }\Phi ^{\delta \omega }\right) \left( 2\Psi _{\alpha \rho
}\Psi _{\beta \sigma }-\Psi _{\alpha \beta }\Psi _{\rho \sigma }\right) \Phi
_{\delta }^{\rho \sigma }\Phi _{\omega }^{\alpha \beta }  \notag \\
&&+8\Phi ^{\rho \sigma }\Psi _{\alpha \beta }\Phi _{\sigma }^{\nu \beta
}\Phi _{\rho }^{\mu \alpha }-8\Phi ^{\mu \alpha }\Psi _{\beta \rho }\Phi
_{\sigma }^{\nu \beta }\Phi _{\alpha }^{\rho \sigma }-8\Phi ^{\nu \alpha
}\Psi _{\beta \rho }\Phi _{\sigma }^{\mu \beta }\Phi _{\alpha }^{\rho \sigma
}  \notag \\
&&+4\Phi ^{\mu \nu }\Psi _{\alpha \rho }\Phi _{\sigma }^{\alpha \beta }\Phi
_{\beta }^{\rho \sigma }\quad .
\end{eqnarray}%
Although the two terms on the right side of (\ref{ebalharm}) may be referred
to loosely as the "field" and "matter" parts of the energy tensor, it should
be remembered that they are not just the functional derivatives of $L_{g}$
and $L_{m}$, as given by (\ref{lgderiv}) and (\ref{lmderiv}); some terms
were transferred from one of these to the other. For most purposes it will
suffice to consider them both together, and then they may be computed as the
left side of (\ref{ebalharm}), that is%
\begin{equation}
t_{H}^{\mu \nu }=\frac{1}{2\kappa }D_{\alpha }D_{\beta }\left( \Phi ^{\mu
\nu }\Phi ^{\alpha \beta }-\Phi ^{\mu \alpha }\Phi ^{\nu \beta }\right)
\quad .
\end{equation}%
At this point we specialize to the case of cartesian harmonic coordinates,
to be discussed in more detail in the next Section, and for which the
Minkowski covariant derivatives are ordinary partial ones. Then the
00-component of $t_{H}^{\mu \nu }$ reduces to%
\begin{equation}
t_{H}^{00}=\frac{1}{2\kappa }\partial _{i}\partial _{j}\left( \Phi ^{00}\Phi
^{ij}-\Phi ^{0i}\Phi ^{0j}\right) \quad ,
\end{equation}%
which may be expressed as a 3-divergence%
\begin{equation}
t_{H}^{00}=\text{div}\mathbf{t,\quad }t_{i}=\frac{1}{2\kappa }\partial
_{j}\Theta _{ij},\quad \Theta _{ij}=\partial _{j}\left( \Phi ^{00}\Phi
^{ij}-\Phi ^{0i}\Phi ^{0j}\right) \quad .  \label{00comp}
\end{equation}

A general spherosymmetric field may be written, in cartesian coordinates $%
x_{0}=t,\sqrt{x_{1}^{2}+x_{2}^{2}+x_{3}^{2}}=r,x_{i}/r=n_{i}$, as%
\begin{equation}
\Phi ^{00}=\Phi _{1}\left( r,t\right) ,\quad \Phi ^{0i}=\Phi _{2}\left(
r,t\right) n_{i},\quad \Phi ^{ij}=\Phi _{3}\left( r,t\right) n_{i}n_{j}+\Phi
_{4}\left( r,t\right) \left( \delta _{ij}-n_{i}n_{j}\right)
\end{equation}%
and its determinant is%
\begin{equation}
\Phi =\left( \Phi _{1}\Phi _{3}-\Phi _{2}^{2}\right) \Phi _{4}^{2}\quad ,
\end{equation}%
and so the cartesian tensor on the right side of (\ref{00comp}) becomes%
\begin{equation}
\Theta _{ij}=\frac{\Phi }{\Phi _{4}^{2}}n_{i}n_{j}+\Phi _{1}\Phi _{4}\left(
\delta _{ij}-n_{i}n_{j}\right) \quad ,  \label{thetaij}
\end{equation}%
that is%
\begin{equation}
\mathbf{t}=\frac{\mathbf{n}}{2\kappa }\left[ \partial _{r}\frac{\Phi }{\Phi
_{4}^{2}}+\frac{2}{r}\left( \frac{\Phi }{\Phi _{4}^{2}}-\Phi _{1}\Phi
_{4}\right) \right] \quad .
\end{equation}

As a first application we use $t_{H}^{00}$ to find the field energy
distribution in the exterior of a nonstatic spherosymmetric $T^{\mu \nu }$
with total mass $M$. By Birkhoff's theorem the field is static, and is the
harmonic version of the Schwarzschild metric, that is (see Ref\cite{fock}
209-215 \ ) putting $m=GM/c^{2}$ 
\begin{equation}
\Phi _{1}=\frac{\left( r+m\right) ^{3}}{r^{2}\left( r-m\right) },\quad \Phi
_{2}=0,\quad \Phi _{3}=-\frac{r^{2}-m^{2}}{r^{2}},\quad \Phi _{4}=-1,\quad
\Phi =-\left( \frac{r+m}{r}\right) ^{4}\quad ,
\end{equation}%
it follows that%
\begin{equation}
\mathbf{t}=\frac{Mc^{2}\mathbf{n}}{4\pi r^{2}}\left( \frac{r+m}{r}\right)
^{3}\frac{2r-m}{2r-2m}\quad .
\end{equation}%
We may define a new vector $\mathbf{t}_{1}$ by subtracting \ a
divergence-free vector%
\begin{equation}
\mathbf{t}_{1}=\mathbf{t-}\frac{Mc^{2}\mathbf{n}}{4\pi r^{2}}\quad ,
\end{equation}%
and then the total (matter plus field) energy contained in $r>r_{1}$ is $%
Mc^{2}\mu \left( r_{1}\right) $, where%
\begin{equation}
\mu \left( r_{1}\right) =\int_{r>r_{1}}\text{div}\mathbf{t}_{1}d^{3}\mathbf{r%
}=1-\left( \frac{r_{1}+m}{r_{1}}\right) ^{3}\frac{2r_{1}-m}{2r_{1}-2m}\quad ,
\label{mufree}
\end{equation}%
but in this case it is entirely field energy. Note that this quantity is
negative; if $r_{1}\gg m$ it is%
\begin{equation}
\int_{r_{1}}^{\infty }t_{H}^{00}d^{3}\mathbf{r}=-\frac{7}{2}Mc^{2}\left[ 
\frac{m}{r_{1}}+O\left( \frac{m^{2}}{r_{1}^{2}}\right) \right] \quad ,
\end{equation}%
and, for example, the absolute value of the field energy in the exterior
space of our Sun, for which $m/r_{1}\approx 2\times 10^{-6},$ is about three
times the mass of our planet.

More generally, we now show that, in the spherosymmetric case, there is a
simple connection between the cartesian tensor $\Theta _{ij}$ and the
spatial components $g_{ij}$, of the Riemannian metric of GR. In terms of the
components of the field tensor we obtained (see equation (\ref{thetaij}))%
\begin{equation}
\Theta _{ij}=\frac{\Phi }{\Phi _{4}^{2}}n_{i}n_{j}+\Phi _{1}\Phi _{4}\left(
\delta _{ij}-n_{i}n_{j}\right) \quad .
\end{equation}%
The inverse field tensor is 
\begin{equation}
\Psi _{00}=\frac{\Phi _{3}\Phi _{4}^{2}}{\Phi },\quad \Psi _{0i}=-\frac{\Phi
_{2}\Phi _{4}^{2}}{\Phi }n_{i},\quad \Psi _{ij}=\frac{\Phi _{1}\Phi _{4}^{2}%
}{\Phi }n_{i}n_{j}+\frac{1}{\Phi _{4}}\left( \delta _{ij}-n_{i}n_{j}\right)
\end{equation}%
Since in these coordinates $\gamma =-1$, the corresponding Riemannian metric
tensor, from (\ref{riemet}), is

\begin{equation}
g_{\alpha \beta }=\sqrt{-\Phi }\Psi _{\alpha \beta }\quad ,
\end{equation}%
that is%
\begin{equation}
g_{00}=\frac{\Phi _{3}\Phi _{4}^{2}}{\sqrt{-\Phi }},\quad g_{0i}=-\frac{\Phi
_{2}\Phi _{4}^{2}}{\sqrt{-\Phi }}n_{i},\quad g_{ij}=\frac{\Phi _{1}\Phi
_{4}^{2}}{\sqrt{-\Phi }}n_{i}n_{j}+\frac{\sqrt{-\Phi }}{\Phi _{4}}\left(
\delta _{ij}-n_{i}n_{j}\right) \quad .
\end{equation}%
Converting to spherical polar coordinates we have the metric%
\begin{equation}
ds^{2}=g_{00}dt^{2}+2g_{0r}dtdr+g_{rr}dr^{2}+g_{\theta \theta }\left(
d\theta ^{2}+\sin ^{2}\theta d\phi ^{2}\right) \quad ,
\end{equation}%
where%
\begin{equation}
g_{0r}=-\frac{\Phi _{2}\Phi _{4}^{2}}{\sqrt{-\Phi }},\quad g_{rr}=\frac{\Phi
_{1}\Phi _{4}^{2}}{\sqrt{-\Phi }},\quad g_{\theta \theta }=r^{2}\frac{\sqrt{%
-\Phi }}{\Phi _{4}}\quad .
\end{equation}%
Then (\ref{thetaij}) becomes%
\begin{equation}
\Theta _{ij}=-\frac{g_{\theta \theta }^{2}}{r^{4}}n_{i}n_{j}-\frac{g_{\theta
\theta }g_{rr}}{r^{2}}\left( \delta _{ij}-n_{i}n_{j}\right) \quad ,
\end{equation}%
which gives%
\begin{equation}
t_{H}^{00}=\text{div}\mathbf{t,\quad t}=-\frac{\mathbf{n}}{\kappa }\left( 
\frac{\partial g_{\theta \theta }}{\partial r}+\frac{g_{\theta \theta
}-g_{rr}}{r}\right) g_{\theta \theta }\quad ,
\end{equation}%
and hence the total energy (matter plus field) in $r>r_{1}$ at time $t$ is $%
\mu \left( r_{1},t\right) Mc^{2},$ where%
\begin{equation}
1-\mu \left( r,t\right) =\frac{g_{\theta \theta }}{2mr^{3}}\left( g_{\theta
\theta }+r^{2}g_{rr}-r\frac{\partial g_{\theta \theta }}{\partial r}\right)
\quad .  \label{murt}
\end{equation}

\section{\protect\bigskip The Oppenheimer-Snyder solution}

We now extend the latter calculation to a continuous mass distribution in
order to discuss the gravitational collapse of a spherical star. In contrast
with the negative energy distribution of the field, the mass tensor\footnote{%
We use the short name, favoured by Fock, for this tensor. Neither it nor the
more usual "material stress tensor" are quite exact according to our present
perspective, because the gravitational field is material and has both energy
and mass.} $T^{\mu \nu }$ has the positive-energy property that its
contraction $T^{\mu \nu }\eta _{\mu }\eta _{\nu }$ with any timelike
covariant vector $\eta _{\mu }$ is positive. One of the few exact solutions
for a Riemannian metric derived from such a mass tensor is that of
Oppenheimer and Snyder\cite{oppsny} (OS) derived in a comoving and
synchronous coordinate system, for which the only nonzero component of $%
T^{\mu \nu }$ is a positive $T^{00}$,%
\begin{equation}
ds^{2}=d\tau ^{2}-\left\{ \partial _{R}^{\ast }W\right\}
^{2}dR^{2}-W^{2}\left( d\theta ^{2}+\sin ^{2}\theta d\phi ^{2}\right) \quad ,
\end{equation}%
where the operator $\partial _{R}^{\ast }$ denotes differentiation with
respect to $R$ at constant $\tau $, and 
\begin{equation}
W=u\left[ \sqrt{\frac{R^{3}}{u^{3}}}-\frac{3\tau \sqrt{2m}}{2}\right]
^{2/3}\quad .
\end{equation}%
The function $u\left( R\right) $ gives the stellar material density, and is
a positive monotone increasing function for $R>0$ with $u\left( 0\right) =0$
and $u\left( \infty \right) =1$. We now choose units such that $2m=1$ and
define a new coordinate 
\begin{equation}
v=\sqrt{\frac{W}{u}}=\left[ \sqrt{\frac{R^{3}}{u^{3}}}-\frac{3\tau }{2}%
\right] ^{1/3}\quad ,
\end{equation}%
to replace $\tau $, so that, denoting differentiation at constant $v$ by $%
\partial _{R}$ , 
\begin{equation}
\partial _{\tau }=-\frac{1}{2v^{2}}\partial _{v},\quad \partial _{R}^{\ast
}=\partial _{R}+\frac{\beta }{2v^{2}}\partial _{v},\quad \beta \left(
R\right) =\left( \frac{R}{u}\right) ^{3/2}\left( \frac{1}{R}-\frac{u^{\prime
}}{u}\right) \quad ,
\end{equation}%
and%
\begin{equation}
\partial _{R}^{\ast }W=\xi v^{2},\quad \xi =u^{\prime }+\frac{\beta u}{2v^{3}%
}\quad .
\end{equation}%
Then the Riemannian interval in terms of $\left( R,v,\theta ,\phi \right) $
is%
\begin{equation}
ds^{2}=\left( \beta dR-2v^{2}dv\right) ^{2}-\xi
^{2}v^{4}dR^{2}-u^{2}v^{4}\left( d\theta ^{2}+\sin ^{2}\theta d\phi
^{2}\right) \quad ,  \label{osmetric}
\end{equation}%
that is%
\begin{eqnarray}
g_{vv} &=&4v^{4},\quad g_{vR}=-2\beta v^{2},\quad g_{RR}=\beta ^{2}-\xi
^{2}v^{4}\quad ,  \notag \\
g_{\theta \theta } &=&\csc ^{2}\theta g_{\phi \phi }=-u^{2}v^{4}\quad .
\end{eqnarray}%
The contravariant inverse \ of this (we are now reverting to conventional
notation) is,%
\begin{eqnarray}
g^{RR} &=&\frac{1}{4v^{4}}-\frac{\beta ^{2}}{4\xi ^{2}v^{8}},\quad g^{vR}=-%
\frac{\beta }{2\xi ^{2}v^{6}},\quad g^{vv}=-\frac{1}{\xi ^{2}v^{4}}\quad , 
\notag \\
g^{\theta \theta } &=&\sin ^{2}\theta g^{\phi \phi }=-\frac{1}{u^{2}v^{4}}%
\quad .
\end{eqnarray}%
The Cartesian harmonic coordinates $x^{\mu }$ are obtained, (see Ref\cite%
{weinberg} pp165-168), as solutions of the equations 
\begin{equation}
\square _{g}x^{\mu }=0\quad \left( \mu =0,1,2,3\right) \quad ,  \label{harm}
\end{equation}%
where $\square _{g}$ is the Riemannian (not to be confused with the
Minkowskian (\ref{minkdal})) d'Alembertian 
\begin{equation}
\square _{g}=\frac{1}{\sqrt{-g}}\partial _{\alpha }\left( g^{\alpha \beta }%
\sqrt{-g}\partial _{\beta }\right) \quad .
\end{equation}%
In spherical coordinates ($t=x_{0},r=\sqrt{x_{1}^{2}+x_{2}^{2}+x_{3}^{2}}$)
these reduce to two equations for what OS term the "exterior time" and
"exterior radius", namely%
\begin{equation}
Qt=0,\quad \left( Q-\frac{2\xi ^{2}}{u^{2}}\right) r=0\quad ,
\label{spherharm}
\end{equation}%
where $Q$ $=-\xi ^{2}v^{4}\square _{g}$ and is given by%
\begin{equation}
Q=\frac{\xi }{u^{2}v^{2}}\left( \partial _{R}+\frac{\beta }{2v^{2}}\partial
_{v}\right) \frac{u^{2}v^{2}}{\xi }\left( \partial _{R}+\frac{\beta }{2v^{2}}%
\partial _{v}\right) -\frac{\xi }{4v^{4}}\partial _{v}v^{4}\xi \partial
_{v}\quad .
\end{equation}%
However, in the present treatment the coordinates $\left( t,r\right) $
describe both the exterior and the interior of the star.

The free-space limit of the last Section corresponds to $u=1,\beta =\sqrt{R}%
,\xi =\sqrt{R}/v^{3}$, for which the solutions of the latter equations are%
\cite{logmest} $\ $ 
\begin{equation}
r=v^{2}-\frac{1}{2},\quad t=\frac{2}{3}R^{3/2}-2\Psi \left( v\right) ,\quad
\Psi \left( v\right) =\int_{w_{0}}^{v}\frac{w^{4}dw}{w^{2}-1}\quad \left(
w_{0}>1\right) \quad ,  \label{freesol}
\end{equation}%
and, substituted in the OS metric (\ref{osmetric}), these give the harmonic
version of the Schwarzschild metric, that is%
\begin{equation}
ds^{2}=\frac{2r-1}{2r+1}dt^{2}-\frac{2r+1}{2r-1}dr^{2}-\left( r+\frac{1}{2}%
\right) ^{2}\left( d\theta ^{2}+\sin ^{2}\theta d\phi ^{2}\right) \quad ,
\end{equation}%
and this corresponds to the far field $\Phi ^{\mu \nu }$ we used in the
previous Section. Both of the PDEs (\ref{spherharm}) have a singularity
along the curve $v=v_{0}\left( R\right) $, specified by%
\begin{equation}
\frac{dv_{0}}{dR}=\frac{1}{2}u^{\prime }-\frac{\beta }{2v_{0}^{3}}\left(
v_{0}-u\right) ,\quad v_{0}\left( \infty \right) =1\quad .  \label{limchar}
\end{equation}%
In our earlier article\cite{cosmo}, we showed, for a particular choice of $u$%
, how to solve (\ref{spherharm}) numerically by integrating along a family
of characteristics, satisfying the ordinary differential equation%
\begin{equation}
\frac{dv}{dR}=\frac{1}{2}u^{\prime }-\frac{\beta }{2v^{3}}\left( v-u\right)
\quad .
\end{equation}%
The operation of differentiation along such a characteristic is%
\begin{equation}
\partial _{R}^{C}=\partial _{R}-\left( \frac{\xi }{2}-\frac{\beta }{2v^{2}}%
\right) \partial _{v}\quad ,
\end{equation}%
and the operator $Q$ may be expressed as%
\begin{equation}
Q=\left( \partial _{R}^{C}\right) ^{2}+\xi \partial _{v}\partial
_{R}^{C}+q\partial _{R}^{C}+\xi ^{2}\left( \frac{1}{u}-\frac{1}{v}\right)
\partial _{v}\quad ,
\end{equation}%
where%
\begin{equation}
q=\frac{2u^{\prime }}{u}+\frac{u^{\prime }}{\xi }+\frac{u}{\xi }\left( \frac{%
5\beta ^{2}}{2v^{6}}-\frac{\beta ^{\prime }}{v^{3}}\right) \quad .
\end{equation}%
We now define the \emph{physical region }as $R>1,v>v_{0}\left( R\right) $;
it is our contention that (\ref{spherharm}) describes the whole evolution of
the collapsing system, without any singularity, in this region, and the
final stage of the collapse, corresponding to the limit $t\left( R,v\right)
\rightarrow +\infty $, is described by the values of $r$ and $t$ close to
the limit characteristic $v=v_{0}\left( R\right) $. A necessary preliminary
to a numerical study of these equations is the obtaining of asymptotic
expansions, both for large $R$ and for large $v$ when $R$ is small, and this
was done in our earlier article\cite{cosmo}. An extension of this
calculation may now be made by applying these earlier results to the total
energy distribution given by (\ref{murt}), putting%
\begin{equation}
g_{\theta \theta }=-u^{2}v^{4}\quad ,
\end{equation}%
and%
\begin{eqnarray}
g_{rr} &=&\left( 2v^{2}\frac{\partial v}{\partial r}-\beta \frac{\partial R}{%
\partial r}\right) ^{2}-\xi ^{2}v^{4}\left( \frac{\partial R}{\partial r}%
\right) ^{2}  \notag \\
&=&\frac{4v^{4}}{J^{2}}\frac{\partial ^{C}t}{\partial R}\left( \frac{%
\partial ^{C}t}{\partial R}+\xi \frac{\partial t}{\partial v}\right) \quad ,
\end{eqnarray}%
where%
\begin{equation}
J=\frac{\partial ^{C}t}{\partial R}\frac{\partial r}{\partial v}-\frac{%
\partial ^{C}r}{\partial R}\frac{\partial t}{\partial v}\quad ,
\end{equation}%
to give%
\begin{equation}
1-\mu \left( R,v\right) =\frac{u^{4}v^{8}}{2mr^{3}}\left[ 1-\frac{2r}{J}%
\left\{ \frac{2}{v}\frac{\partial ^{C}t}{\partial R}-\xi \left( \frac{1}{u}-%
\frac{1}{v}\right) \frac{\partial t}{\partial v}\right\} -\frac{4r^{2}}{%
J^{2}u^{2}}\frac{\partial ^{C}t}{\partial R}\left( \frac{\partial ^{C}t}{%
\partial R}+\xi \frac{\partial t}{\partial v}\right) \right] 
\end{equation}%
.

It is appropriate to mention again here that $t_{H}^{\mu \nu }\left(
r,t\right) ,$ from which the quantity $\mu \left( R,v\right) $ was derived,
is indeed a tensor. If, for example, we should choose to transform it back
to the original OS coordinates, this would be possible, provided we bear in
mind that in those coordinates the partial derivatives $\partial _{\alpha }$
and $\partial _{\beta }$ would be replaced by the Minkowski covariant
derivatives $D_{\alpha }$ and $D_{\beta }$; the latter, of course, are not
the same as the Riemann covariant derivatives $\nabla _{\alpha }$ and $%
\nabla _{\beta }$. In order to describe the actual energy distribution we
need $\mu $ as a function of $\left( r,t\right) $, which requires rather
complicated interpolations from $\mu \left( R,v\right) ,r\left( R,v\right) $
and $t\left( R,v\right) ,$ and we shall describe such a detailed calculation
in another article. Nevertheless we venture to draw some qualitative
conclusions from the results already obtained.

\section{\protect\bigskip What we expect to find}

The field part of the 00-component of the energy tensor is negative, as we
found in an earlier section when we examined the far field. In the deep
interior of the star we expect to find that this component is substantial,
and indeed bigger in absolute magnitude than $T^{00}$. Being negative, it
produces, in a bootstrapping manner, its own gravitational field, \emph{and
this is repulsive}. That explains why, as we already found in our earlier
article\cite{cosmo}, the material energy $T^{00}$ is concentrated in a shell
near the surface as $t\rightarrow +\infty $. The density function $u\left(
R\right) $ describes a star with an initially diffuse corona, but the
formation of a distinct surface is an expected part of its evolution. As
pointed out above, a substantial amount of computation is required in order
to obtain a quantitative expression for the total (gravitational energy plus
stellar material) mass distribution, $\mu \left( r,t\right) .$ In Figure 1
we give an artist's impression of this profile, for fixed $t$, at a fairly
advanced stage of the collapse process, when the surface shell has begun to
form.\FRAME{ftbpFU}{4.1597in}{3.2776in}{0pt}{\Qcb{The energy content $1-%
\protect\mu \left( r\right) $, in units of $Mc^{2}$, contained within a
sphere of radius $r$ measured from the centre of a collapsing star, as a
function of $r$ for fixed $t$. Between A and B the energy is mainly
gravitational and negative, between B and C it is mainly stellar material
and positive, and beyond C it is again gravitational and negative. The
radius $r$ is in units of the Schwarzschild radius.}}{}{mnras.eps}{\special%
{language "Scientific Word";type "GRAPHIC";maintain-aspect-ratio
TRUE;display "USEDEF";valid_file "F";width 4.1597in;height 3.2776in;depth
0pt;original-width 6.6789in;original-height 5.2581in;cropleft "0";croptop
"1";cropright "1";cropbottom "0";filename '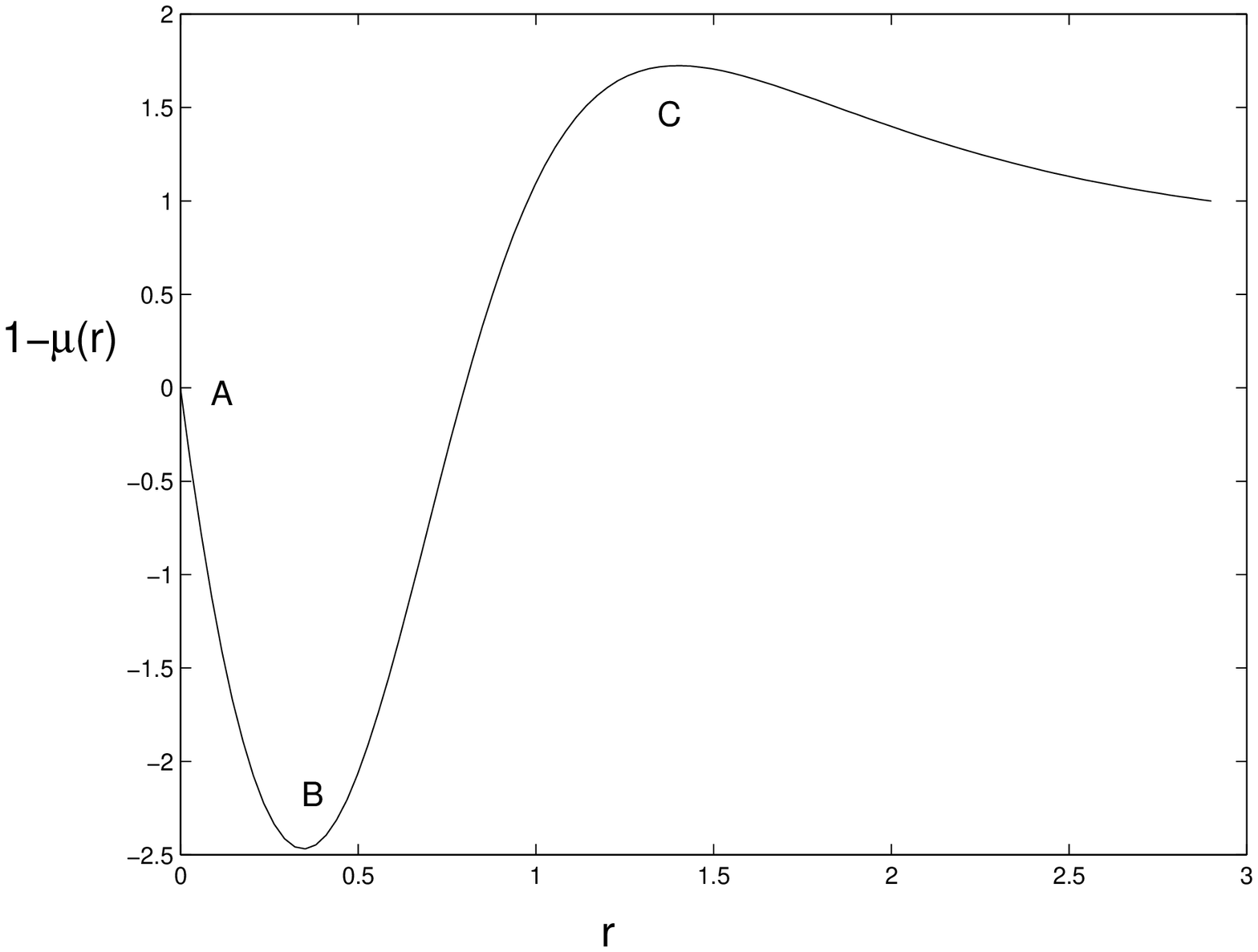';file-properties
"XNPEU";}}

The key element in our approach, stressed also in our earlier article\cite%
{cosmo}, is that the coordinate-free topological analysis, now common in GR,
fails to acknowledge the central role of the energy tensor. We accept the
Hilbert-Einstein field equations, but add a supplementary set which results
in the inertial (that is harmonic) coordinate system being privileged. That
means maintaining both covariance and a weak principle of equivalence, in
the form of EoP, but rejecting gauge invariance. As was emphasized first by
Fock, this latter "loss" means taking the "G" out of GR, leaving us,
nevertheless, with a theory which Fock himself called Einstein's Theory of
Gravitation.

As far as the currently fashionable notion of "black holes" is concerned,
our conclusion is that there is a limit to the degree of compactification
undergone in a collapsing gravitational system, and this limit is set by the
properties of the gravitational field itself rather than the presence of the
other forces of nature, like neutron degenerative pressure. Relatively small
"condensars" are observed as neutron stars, or pulsars, but heavier ones,
like the one thought to lie at the centre of our galaxy, are more diffuse,
probably of white-dwarf density. Our analysis is of an idealized, purely
gravitational system, and our key finding is the limit characteristic curve,
which represents the limit $t\rightarrow +\infty .$ This curve is equivalent
to the "event horizon" popularly acknowledged to come out of coordinate-free
GR, but for us there is \emph{nothing} the other side of the limit
characteristic. The other side of our "event horizon" lies outside the
physical space; to go there is to go "to infinity and beyond" along with
maybe Buzz Lightyear or Doctor Who!


\begin{thebibliography}{99}
\bibitem{wheeler} C. \ W. Misner, K. S. Thorne and J. A. Wheeler, \emph{%
Gravitation, }(Freeman, San Francisco, 1973)

\bibitem{Einstein1} A. Einstein, \emph{Sitzungsber. preuss. Akad. Wiss. }%
\textbf{48, }844-847 (1915)

\bibitem{Einstein2} A. Einstein, \emph{Naturforsch. Gesellschaft Zurich }%
\textbf{58, }284-290 (1913)

\bibitem{Einstein3} A. Einstein, \emph{Sitzungsber. preuss. Akad. Wiss. }%
\textbf{1, }154-167 (1918)

\bibitem{Kennefick} D. Kennefick,\emph{Traveling at the Speed of Thought }%
(Princeton Univ Press, 2007)

\bibitem{HulseT} R. A. Hulse and J. H. Taylor, \emph{Astrophys. J. }\textbf{%
195, }L51-53 (1975)

\bibitem{LISA} \emph{Report on the LIGO project, }%
http://www.ligo-la.caltech.edu/LLO/overviewsci.htm

\bibitem{Eddington} A. S. Eddington, \emph{The Mathematical Theory of
Relativity, }(Cambridge Univ Press,1923)

\bibitem{Hilbert} D. Hilbert, \emph{Goettinger Nachrichten }\textbf{4},21
(1917)

\bibitem{dedond} T. de Donder,\emph{\ La gravitique einsteinienne},
(Gauthier Villars, Paris, 1921)

\bibitem{lanczos} C. Lanczos, \emph{Phys. Z. }\textbf{23}, 537 (1923)

\bibitem{fock} V. A. Fock, \emph{Space, Time and Gravitation, }(Pergamon,
Oxford,1966)

\bibitem{rosen} N. Rosen, \emph{Phys. Rev. }\textbf{57, }147-153 (1940)

\bibitem{weinberg} S. Weinberg, \emph{Gravitation and Cosmology }(John
Wiley, New York, 1972)

\bibitem{logmest} A. A. Logunov and M. A. Mestvirishvili, \emph{The
Relativistic Theory of Gravitation }(Mir, Moscow, 1989)

\bibitem{babak} S. V. Babak and L. P. Grishchuk, \emph{Phys. Rev. D }\textbf{%
61, }024038 (1999)

\bibitem{oppsny} J. R. Oppenheimer and H. Snyder, \emph{Phys.Rev. }\textbf{%
56, }455-459 (1939)

\bibitem{LL} L. D. Landau and E. M. Lifshitz, \emph{Classical Theory of
Fields }(Addison Wesley, Cambridge, Mass, 1951)

\bibitem{logunov} A. A. Logunov, \emph{Theory of Gravity }(Nauka, Moscow,
2001)

\bibitem{cosmo} T. W. Marshall and M. K. Wallis \emph{J. Cosmology, }\textbf{%
6, }1473-1484 (2010)
\end{thebibliography}
\end{document}